\input harvmac
%
\ifx\epsfbox\UnDeFiNeD\message{(NO epsf.tex, FIGURES WILL BE
IGNORED)}
\def\figin#1{\vskip2in}
\else\message{(FIGURES WILL BE INCLUDED)}\def\figin#1{#1}\fi
\def\ifig#1#2#3{\xdef#1{fig.~\the\figno}
\goodbreak\midinsert\figin{\centerl ine{#3}}%
\smallskip\centerline{\vbox{\baselineskip12pt
\advance\hsize by -1truein\noindent\footnotefont{\bf
Fig.~\the\figno:} #2}}
\bigskip\endinsert\global\advance\figno by1}

\def\footnotefont{\tenpoint}

\newwrite\ffile\global\newcount\figno \global\figno=1
\def\fig{fig.~\the\figno\nfig}
\def\nfig#1{\xdef#1{fig.~\the\figno}%
\writedef{#1\leftbracket fig.\noexpand~\the\figno}%
\ifnum\figno=1\immediate\openout\ffile=figs.tmp\fi\chardef\wfile=\ffile%
\immediate\write\ffile{\noexpand\medskip\noexpand\item{Fig.\
\the\figno. }
\reflabeL{#1\hskip.55in}\pctsign}\global\advance\figno
by1\findarg}

\overfullrule=0pt
\parindent 25pt
\tolerance=10000
\sequentialequations

\def\lr{\lref}

\lr\beckera{
K.~Becker, M.~Becker and A.~Strominger,
``Five-branes, membranes and nonperturbative string theory,''
Nucl.\ Phys.\  {\bf B456} (1995) 130
[hep-th/9507158].}
\lr\beckerb{K.~Becker and M.~Becker,
``Instanton action for type II hypermultiplets,''
Nucl.\ Phys.\  {\bf B551} (1999) 102
[hep-th/9901126].}
\lr\stromingera{A.~Strominger,
``Loop corrections to the universal hypermultiplet,''
Phys.\ Lett.\  {\bf B421} (1998) 139
[hep-th/9706195].}
\lr\bagger{J.~Bagger and E.~Witten,
``Matter Couplings In N=2 Supergravity ,''
Nucl.\ Phys.\  {\bf B222} (1983) 1.}
\lr\ferraraa{S.~Cecotti, S.~Ferrara and L.~Girardello,
``Geometry Of Type II Superstrings And The Moduli Of Superconformal Field Theories,''
Int.\ J.\ Mod.\ Phys.\  {\bf A4} (1989) 2475.}
\lr\ferrarab{S.~Ferrara and S.~Sabharwal,
``Quaternionic Manifolds For Type II Superstring Vacua Of Calabi-Yau Spaces,''
Nucl.\ Phys.\  {\bf B332} (1990) 317.}
\lr\ferrarac{I.~Antoniadis, S.~Ferrara, R.~Minasian and K.~S.~Narain,
``R**4 couplings in M- and type II theories on Calabi-Yau spaces,''
Nucl.\ Phys.\  {\bf B507} (1997) 571}
\lr\louisa{H.~Gunther, C.~Herrmann and J.~Louis,
``Quantum corrections in the hypermultiplet moduli space,''
Fortsch.\ Phys.\  {\bf 48} (2000) 119
[hep-th/9901137].}
\lr\ooguria{H.~Ooguri, Y.~Oz and Z.~Yin,
``D-branes on Calabi-Yau spaces and their mirrors,''
Nucl.\ Phys.\  {\bf B477} (1996) 407
[hep-th/9606112].}
\lr\gutperlea{M.~Gutperle and Y.~Satoh,
``D-branes in Gepner models and supersymmetry,''
Nucl.\ Phys.\  {\bf B543} (1999) 73
[hep-th/9808080].}
\lr\greena{M.~B.~Green and M.~Gutperle,
``Effects of D-instantons,''
Nucl.\ Phys.\  {\bf B498} (1997) 195
[hep-th/9701093].}
\lr\harveya{J.~A.~Harvey and G.~Moore,
``Fivebrane instantons and R**2 couplings in N = 4 string theory,''
Phys.\ Rev.\  {\bf D57} (1998) 2323
[hep-th/9610237].}
\lr\kiritsisa{C.~Bachas, C.~Fabre, E.~Kiritsis, N.~A.~Obers and P.~Vanhove,
``Heterotic/type-I duality and D-brane instantons,''
Nucl.\ Phys.\  {\bf B509} (1998) 33
[hep-th/9707126]. }
\lr\kiritsisb{B.~Pioline and E.~Kiritsis,
``U-duality and D-brane combinatorics,''
Phys.\ Lett.\  {\bf B418} (1998) 61
[hep-th/9710078].}
\lr\kiritsisc{E.~Kiritsis, N.~A.~Obers and B.~Pioline,
``Heterotic/type II triality and instantons on K3,''
JHEP {\bf 0001} (2000) 029
[hep-th/0001083].}
\lr\hammou{A.~B.~Hammou and J.~F.~Morales,
``Fivebrane instantons and higher derivative couplings in type I theory,''
hep-th/9910144.}
\lr\harveyb{J.~A.~Harvey and G.~Moore,
``Superpotentials and membrane instantons,''
hep-th/9907026.}
\lr\cremmer{E.~Cremmer, B.~Julia and J.~Scherk,
``Supergravity theory in 11 dimensions,''
Phys.\ Lett.\  {\bf B76} (1978) 409.}
\lr\gibbonsa{G.~W.~Gibbons, M.~B.~Green and M.~J.~Perry,
``Instantons and Seven-Branes in Type IIB Superstring Theory,''
Phys.\ Lett.\  {\bf B370}, 37 (1996)
[hep-th/9511080].}
\lr\lusta{K.~Behrndt, I.~Gaida, D.~Lust, S.~Mahapatra and T.~Mohaupt,
``From type IIA black holes to T-dual type IIB D-instantons in N = 2,  D = 4 supergravity,''
Nucl.\ Phys.\  {\bf B508} (1997) 659
[hep-th/9706096].}
\lr\stellea{E.~Cremmer, I.~V.~Lavrinenko, H.~Lu, C.~N.~Pope, K.~S.~Stelle and T.~A.~Tran,
``Euclidean-signature supergravities, dualities and instantons,''
Nucl.\ Phys.\  {\bf B534}, 40 (1998)
[hep-th/9803259].}
\lr\cadavid{A.~C.~Cadavid, A.~Ceresole, R.~D'Auria and S.~Ferrara,
``Eleven-dimensional supergravity compactified on Calabi-Yau threefolds,''
Phys.\ Lett.\  {\bf B357} (1995) 76
[hep-th/9506144].}
\lr\ovruta{A.~Lukas, B.~A.~Ovrut, K.~S.~Stelle and D.~Waldram,
``Heterotic M-theory in five dimensions,''
Nucl.\ Phys.\  {\bf B552} (1999) 246
[hep-th/9806051].}
\lr\coleman{S.~Coleman and K.~Lee,
``Wormholes Made Without Massless Matter Fields,''
Nucl.\ Phys.\  {\bf B329} (1990) 387.}
\lr\stromingerc{S.~B.~Giddings and A.~Strominger,
``String Wormholes,''
Phys.\ Lett.\  {\bf B230} (1989) 46.}
\lr\waldron{P.~van Nieuwenhuizen and A.~Waldron,
``On Euclidean spinors and Wick rotations,''
Phys.\ Lett.\  {\bf B389}, 29 (1996) [hep-th/9608174].}
\lr\rey{S.~Rey,
``The Confining Phase of Superstrings and Axionic Strings,''
Phys.\ Rev.\  {\bf D43} (1991) 526.}
\lr\towna{J.~M.~Izquierdo, N.~D.~Lambert, G.~Papadopoulos and P.~K.~Townsend,
``Dyonic Membranes,''
Nucl.\ Phys.\  {\bf B460} (1996) 560
[hep-th/9508177].}
\lr\townb{M.~B.~Green, N.~D.~Lambert, G.~Papadopoulos and P.~K.~Townsend,
``Dyonic p-branes from self-dual (p+1)-branes,''
Phys.\ Lett.\  {\bf B384} (1996) 86
[hep-th/9605146].}
\lr\harveyb{J.~A.~Harvey and G.~Moore,
``Superpotentials and membrane instantons,''
hep-th/9907026.}
\lr\aspinwall{P.~S.~Aspinwall,
``Compactification, geometry and duality: N = 2,''
hep-th/0001001.}
\lr\townd{D.~Sorokin and P.~K.~Townsend,
``M-theory superalgebra from the M-5-brane,''
Phys.\ Lett.\  {\bf B412} (1997) 265
[hep-th/9708003].}
\lr\pasti{I.~Bandos, K.~Lechner, A.~Nurmagambetov, P.~Pasti, D.~Sorokin and M.~Tonin,
``Covariant action for the super-five-brane of M-theory,''
Phys.\ Rev.\ Lett.\  {\bf 78} (1997) 4332
[hep-th/9701149].}
\lr\schwarzd{
M.~Aganagic, J.~Park, C.~Popescu and J.~H.~Schwarz,
``World-volume action of the M-theory five-brane,''
Nucl.\ Phys.\  {\bf B496} (1997) 191
[hep-th/9701166].}
\lr\westb{O.~Barwald, N.~D.~Lambert and P.~C.~West,
``A calibration bound for the M-theory fivebrane,''
Phys.\ Lett.\  {\bf B463} (1999) 33
[hep-th/9907170].}
\lr\lustc{D.~Lust and A.~Miemiec,
``Supersymmetric M5-branes with H-field,''
Phys.\ Lett.\  {\bf B476} (2000) 395
[hep-th/9912065].}
\lr\stromingerd{M.~Marino, R.~Minasian, G.~Moore and A.~Strominger,
``Nonlinear instantons from supersymmetric p-branes,''
JHEP {\bf 0001} (2000) 005
[hep-th/9911206].}
\lr\stieberger{W.~Lerche and S.~Stieberger,
``1/4 BPS states and non-perturbative couplings in N = 4 string theories,''
hep-th/9907133.}
\lr\bobby{B.~S.~Acharya,
``M theory, Joyce orbifolds and super Yang-Mills,''
 ATMP {\bf 3} (1999) 227 [hep-th/9812205].}
\noblackbox
\baselineskip 20pt plus 2pt minus 2pt
\Title{\vbox{\baselineskip12pt \hbox{hep-th/0005068}
\hbox{HUTP-00/A013}  }}
{\vbox{\centerline{Supergravity Instantons and the   Universal
Hypermultiplet}}}  

\centerline{Michael Gutperle and Micha\l\ Spali\'nski}
\smallskip
\centerline{\it Jefferson Laboratory of Physics}
\centerline{\it Harvard University, Cambridge, MA 02138, USA}
\bigskip
\centerline{\tt gutperle@riemann.harvard.edu,$\;$ mspal@schwinger.harvard.edu}
\bigskip

\medskip
\centerline{{\bf Abstract}}

The effective action of $N=2$ supersymmetric $5$-dimensional supergravity
arising from compactifications of M-theory on Calabi-Yau threefolds
receives non-perturbative corrections from wrapped Euclidean membranes and 
fivebranes. These contributions can be interpreted as instanton
corrections in the $5$ dimensional field theory. Focusing on the universal
hypermultiplet, a solution of this type is 
presented and the instanton action is calculated, generalizing previous
results involving membrane instantons. The instanton action is not a sum of
membrane and fivebrane contributions: it has the form reminiscent of
non-threshold bound states.

\noblackbox
\baselineskip 20pt plus 2pt minus 2pt

\Date{May 2000}


\newsec{Introduction}

Nonperturbative effects in compactified  string theories and M-theory  can
often be understood in terms of Euclidean wrapped  branes. The study of
such effects for M-theory and  type II string theory was initiated by
Becker, Becker and 
Strominger \beckera. In the case of M-theory on a Calabi-Yau there are
two types of effects: those arising from
membranes wrapping special Lagrangian submanifolds and those from
fivebranes wrapping the whole Calabi-Yau.

The simplest setting for a treatment of these effect is given by
focusing exclusively on the universal hypermultiplet. This multiplet
(leaving aside global issues \aspinwall) is independent of the detailed
structure of the Calabi-Yau. For 
M-theory and type IIA string theory compactified on 
rigid Calabi-Yau manifolds (with $h_{2,1}=0$) the universal hypermultiplet
is the only hypermultiplet. 

Hypermultiplets of $N=2$ supergravity in four and five dimensions
parameterize  quaternionic manifolds \bagger. Classically the universal 
hypermultiplet lives in a $SU(2,1)/U(2)$ coset \ferraraa\ferrarab.
Some aspects of  perturbative corrections to the universal
hypermultiplet obtained by dimensional reduction of higher  derivative
terms in M-theory  were discussed in \stromingera\ferrarac\louisa.   On the
other hand membrane and fivebrane instantons \beckera\ will provide
nonperturbative 
corrections to the metric on the moduli space of the hypermultiplets in the
five dimensional supergravity. 
The study of such effects was
continued in \beckerb, which in particular investigated charge
quantization and the breaking of 
continuous isometries of the quaternionic manifold due to instantons.

After compactification on a circle M-theory reduces to IIA superstring
theory and the the M5 brane and M2 brane instanton will become a NS-5 brane
and D2 brane instanton respectively. Instantons corresponding to wrapped
D-branes can also be described using boundary states \ooguria\ and their
effects can be analyzed \gutperlea\ aanalogously to D-instantons in ten
dimensional IIB theory \greena. 

Membrane instantons in compactifications with more supersymmetry were
discussed \harveya\kiritsisa\kiritsisb\kiritsisc\hammou\stieberger. Membrane
instantons in manifolds of exceptional holonomy (and $N=1$ supersymmetry)
were discussed in \harveyb\bobby.

Solutions of ten dimensional IIB supergravity corresponding to D-instantons
were first obtained in \gibbonsa, see also \lusta\stellea\rey. These instantons
carry ``charges'' associated with shifts of pseudoscalar fields. 
In the case of
the universal hypermultiplet there are three pseudoscalars
(in the four 
dimensional language they are the NS-NS axion and two RR-scalars). 

In this paper the $N=2$ supergravity arising from dimensional reduction
of eleven dimensional  
supergravity \cremmer\ on a Calabi-Yau threefold is studied with the
intention of learning about nonperturbative corrections to the
hypermultiplet moduli space arising from membranes and fivebranes wrapping
Calabi-Yau cycles. 
Instanton solutions are found which carry all three of the charges which
descend from 
membrane and fivebrane charges in $11$ dimensions. By adding appropriate
boundary terms to the Euclidean action the instanton action is
evaluated. An interesting aspect of the result is that it is not a sum of
membrane and fivebrane contributions, but has the form characteristic of a
non-threshold bound state. 

Section 2 reviews compactification of $11$ dimensional supergravity and
exhibits the quaternionic geometry and the isometries relevant to
subsequent considerations. Section 3 discusses the charges. Supersymmetry
preservation conditions are analyzed in section 4. The instanton solution
is presented in section 5. Section 6 is devoted to the evaluation of the
instanton action. Section 7 describes the broken and unbroken
supersymmetries and offers some remarks on computing the fermion
determinant. Conclusions and a discussion of the results is given in
section 8.

\newsec{Eleven dimensional supergravity on a Calabi-Yau manifold}

This section reviews the reduction of eleven dimensional supergravity on a 
Calabi-Yau threefold. 
The bosonic part of the action of eleven dimensional supergravity
\cremmer\ is given by
\eqn\elevsuga{
S= {1\over 2 k_{11}^2}\int d^{11}x \sqrt{-g}\Big( R-{1\over
48}F^{MNPQ}F_{MNPQ}\big) -{1\over 12k_{11}^2} \int A\wedge F\wedge F \ .
}
The supersymmetry transformation of the gravitino in eleven dimensional
supergravity is 
\eqn\susytraf{
\delta_\epsilon\psi_M= \partial_M \epsilon +{1\over
4}\omega_{M}^{\underline A\underline B} \Gamma_{\underline A\underline B}
\epsilon -{1\over 288}\Big( \Gamma_{M}^{NPQR}-8\delta_M^N\Gamma^{PQR}\Big)
\epsilon F_{NPQR}\ .
} 
The notation here is that $\underline A,\underline B$ denote tangent space
indices and $M,N$ denote world indices. 

Dimensional reduction of eleven dimensional supergravity on a Calabi-Yau
manifold (with vanishing G-fluxes), produces five dimensional $N=2$
supergravity with $h_{1,1}-1$ vectormultiplets and $h_{2,1}+1$ 
hypermultiplets \cadavid. The rest of the paper will focus on 
the dynamics of 
the universal hypermultiplet, which is present in any Calabi-Yau
compactification. 

The coordinates are split according to $x^M=(x^\mu,y^m)$, with
$x^\mu,\mu=1,\cdots,6$ parameterizing the internal Calabi-Yau space and
$y^m,m=0,\cdots,4$ parameterizing the transverse space. 
The ansatz for the metric is 
\eqn\metric{
ds^2= e^{-1/3\sigma(y^2)}ds^2_{CY}+ e^{2/3
\sigma(y^2)}(dy_m)^2 \ ,
}
where $ds^2_{CY}$ is an unspecified Ricci flat metric of the Calabi-Yau 
manifold and $y^2\equiv y^m y^m$. 
The other fields in the universal hypermultiplet are the three form
potential $C_{mnk}$ (which is dual to a scalar) 
and two real scalar fields $\chi_1$, $\chi_2$ defined by
\eqn\scalarred{
C_{\alpha\beta\gamma}={1\over \sqrt{2}} \chi
\Omega_{\alpha\beta\gamma},\quad 
C_{\bar\alpha\bar\beta\bar\gamma}={1\over \sqrt{2}}  \bar\chi
\bar\Omega_{\bar\alpha\bar\beta\bar\gamma} \ ,
}
where $\chi=\chi_1+i\chi_2$ and $\bar\chi= \chi_1-i\chi_2$. Here $\Omega$,
$\bar\Omega$ are the unique harmonic $(3,0)$ and $(0,3)$ forms
on the Calabi-Yau manifold. 

Using \metric\ and \scalarred, the bosonic part of the action for the
universal hypermultiplet is given by
\eqn\fivdac{
\eqalign{S&= -{1\over 2 \kappa_5^2}\int d^5x \sqrt{g}\Big(  \partial_m
\sigma\partial_m \sigma +{1\over
24}e^{-2\sigma}F_{mnpq}F^{mnpq}+e^\sigma
\partial_m \chi\partial_m \bar\chi\Big)\cr
&+{i\over
48 \kappa_5^2}\int d^5x \;\epsilon^{mnpql}F_{mnpq}(\bar\chi\partial_l
\chi-\chi\partial_l\bar\chi) \ .} 
}
In the following we set the five dimensional Newton constant
$\kappa_5=1$ for notational convenience.
In order to exhibit the quaternionic structure of the hypermultiplet moduli
space the
four form field strength  has to be dualized. This is accomplished by
introducing a Lagrange multiplier field $D$ and modifying the action as
follows: 
\eqn\lprim{
S^\prime = S- {1\over 24}\int d^5x\;\epsilon^{mnpql}F_{mnpq}
\partial_l D \ . 
}
Integrating out $D$ enforces the Bianchi identity $dF=0$ and one recovers
the original action \fivdac. If, on the other 
hand, one integrates out $F$ the dual description is obtained. The equation of
motion for $F$ is given by
\eqn\eqofmF{
e^{-2\sigma} F^{mnpq}+{i\over 4}\epsilon^{mnpql}(\bar\chi\partial_l
\chi-\chi\partial_l\bar\chi)  -\epsilon^{mnpql}\partial_l D=0 \ .
}
Using this relation the dual action becomes 
\eqn\ldual{
\eqalign{S'&= -\int d^5x\;\Big({1\over 2} \partial_m
\sigma\partial_m \sigma +{1\over 2}e^\sigma
\partial_m\chi\partial_m\bar\chi
+{1\over 2}e^{2\sigma}\big(\partial_l D -{i\over 4}(\bar\chi\partial_l 
\chi-\chi\partial_l\bar\chi)\big)^2\Big) \ . }
}
In this dualized form the four scalars in the universal hypermultiplet
parameterize the 
quaternionic manifold $SU(2,1)/U(2)$ \ferrarab\stromingera. There are three
isometries of this space, which are associated with shifts of the axionic
fields $\chi,\bar\chi$ and $D$\ferrarab\stromingera\beckerb:
\eqn\shift{
\chi\to \chi+\epsilon,\quad
\bar\chi\to\bar\chi+\bar\epsilon,\quad D\to
D+\delta+{i\over 4}\big(\chi\bar\epsilon-\bar\chi \epsilon\big) \ , 
}
with constant parameters $\epsilon=\epsilon_1+i\epsilon_2$
and $\delta$. These isometries are broken to a discrete subgroup by
instanton effects \beckerb.

\newsec{Charges}

Solutions of the supergravity equations of motion may carry charges which
descend from the 
fivebrane and membrane charges of M-theory. 
The solutions of interest here are characterized by three charges
associated with 
shifts of $D$, $\chi$ and $\bar\chi$: these correspond 
to fivebranes wrapped on the entire Calabi-Yau manifold, and to
membranes wrapping $3$-cycles in the Calabi-Yau manifold respectively.
The topological charge associated with the fivebrane is given by 
\eqn\chargefiv{
Q_5 = \oint_{S^4_\infty} F.}
The two membrane charges are the Noether charges associated with
shifts of $\chi$ and $\bar\chi$:
\eqn\twobch{
\eqalign{Q_2&= \oint_{S^4_\infty}(e^\sigma *d \chi-
i \chi F)= q_2+ i Q_5\chi_\infty \ ,\cr
\bar Q_2&= \oint_{S^4_\infty}(e^\sigma *d \bar\chi + i \bar\chi
F)= \bar q_2- i Q_5\bar\chi_\infty \ ,} 
}
where $\chi_\infty$ and $\bar\chi_\infty$ are the asymptotic values (at 
$r=\infty$) of $\chi$ and $\bar\chi$ respectively. 
Unlike the Noether charges $Q_2,\bar
Q_2$ the charges $q_2,\bar q_2$ are invariant under 
shifts of $\chi_\infty$ and $\bar\chi_\infty$. This  behavior can be traced to the
presence of the Chern-Simons 
term in the eleven dimensional supergravity action \elevsuga.

\newsec{Supersymmetry}

The focus of interest here 
are instanton solutions which preserve 
four of the eight supersymmetries of the $d=5$ $N=2$ supergravity. 
The supersymmetry transformations of the five dimensional fields can easily
be derived from the eleven dimensional ones \susytraf\ (see
e.g. \ovruta). The hyperino transformations  are given
by 
\eqn\susytrafh{
\eqalign{\delta \xi^1 &= {1\over 3}\big(\partial_n \sigma
\gamma^n- {i\over 
24} e^{-\sigma}\epsilon^{mnpqr}F_{mnpq}\gamma_r\big)\epsilon^1+
{1\over 3}
e^{\sigma/2 }\partial_n \chi \gamma^n \epsilon^2 \ , \cr
\delta \xi^2 &= -{1\over 3}
e^{\sigma/2 }\partial_n \bar \chi \gamma^n \epsilon^1+{1\over
3}\big(\partial_n \sigma \gamma^n+ {i\over 
24} e^{-\sigma}\epsilon^{mnpqr}F_{mnpq}\gamma_r\big)\epsilon^2 \ .}
}
These transformations can be succinctly expressed in terms of a two-by-two
matrix $M$ (with gamma matrix entries)
as $\delta
\xi^i=M^i_{\;j}\epsilon^j$. Supersymmetry transformations for the gravitini  
are given by  
\eqn\susytrafg{
\eqalign{\delta \psi^1_m &= \big(\nabla_m + {i\over
288} e^{-\sigma}\epsilon_m^{\;\;npqr}F_{npqr}\big)\epsilon_1-
{1\over 3}
e^{\sigma/2 }\partial_m \chi \epsilon_2 \ ,\cr
\delta \psi_m^2 &= {1\over 3}
e^{\sigma/2 }\partial_m \bar \chi  \epsilon_1+\big(\nabla_m - {i\over
288} e^{-\sigma}\epsilon_m^{\;\;npqr}F_{npqr}\big)\epsilon_2 \ .}
}
Supersymmetry transformations for the dualized action \ldual\ are  given
by eliminating $F$ in \susytrafh\susytrafg\ using the  relation
\eqofmF. They then take on the standard form for $N=2$ hypermultiplets
\bagger.

\newsec{Instanton solution}

An instanton in a supergravity theory is a solution to the Euclidean
equations of motion which is localized in the five dimensional
(Euclidean) space time. It carries charges associated with shift
symmetries 
of (axionic) scalars. There is an important sublety concerning the question
how these axionic
scalars are treated when the action is continued from Minkowski to
Euclidean signature. The point of view assumed here is that one   
multiplies the (real) pseudoscalar fields by a factor of $i$ (this changes the sign of
the kinetic term for $\chi,\bar\chi$). Note that this prescription
seemingly makes the real part of the instanton action unbounded from below; this is however rectified by the boundary term
introduced in section 6\foot{See 
\coleman\stromingerc\rey\gibbonsa\waldron\ for a detailed discussion of
this issue.}.

The action \fivdac\ then becomes 
\eqn\fivdace{
\eqalign{S_{eucl}&= \int d^5x \sqrt{g}\Big( {1\over 2} \partial_m
\sigma\partial_m \sigma +{1\over
48}e^{-2\sigma}F_{mnpq}F^{mnpq}-{1\over 2}e^\sigma
\partial_m \chi\partial_m \bar\chi\Big)\cr
&-{1\over
96}\int d^5x \;\epsilon^{mnpql}F_{mnpq}(\bar\chi\partial_l
\chi-\chi\partial_l\bar\chi)\ . }
}
An important simplifying assumption is that the solution is
taken to be rotationally
symmetric in Euclidean spacetime, i.e. all fields are assumed
to depend only on $r\equiv |y-y_0|$, where $y_0$ is the location of the
instanton.  

The required solution is characterized by three charges associated with
shifts of $D$, $\chi$ and $\bar\chi$ (as discussed in section 3): these
correspond  
to fivebranes wrapped on the entire Calabi-Yau manifold, and to
membranes wrapping $3$-cycles in the Calabi-Yau manifold.

The instanton solution should be BPS, i.e. it should preserve half of
the eight supersymmetries parameterized by 
two spinors $\epsilon_{1},\epsilon_2$. For 
rotationally  invariant field configurations  the 
condition that $\delta\xi_i=0$ in \susytrafh\ can be reduced to a two by
two matrix equation. The BPS condition is then equivalent to the 
condition that this matrix has  a zero eigenvalue. This implies,
\eqn\zeroev{
\partial_m\sigma\partial_m\sigma -{1\over
24}e^{-2\sigma}F_{mnpq}F^{mnpq}-
e^{\sigma}\partial_m\chi\partial_m\bar\chi=0 \ . 
}   

The Ansatz for the metric is particularly simple: in the Einstein frame,
the metric is flat $g_{mn}=\delta_{mn}$. In addition the following
Ansatz for the $4$-form field strength is made: 
\eqn\ansatzD{
F_{mnpq}= \epsilon_{mnpqr}\partial_r H \ .
}
The Bianchi identity for $F$ then implies that $H$ is harmonic, hence 
\eqn\etoc{
H =  {1 \over 8\pi^2} {Q_5\over r^3} \ .
}
The normalization here has been chosen so that the parameter $Q_5$
appearing above is the fivebrane
charge, as defined in the previous section.

Using \ansatzD\ the  equation of motion for $\sigma$ derived from the
action  \fivdace\ is given by 
\eqn\eoma{
\partial_m\partial_m \sigma + {1\over 2}e^{\sigma} \partial_m \chi
\partial_m \bar \chi 
 +  e^{-2\sigma} \partial_m H \partial_m H = 0 \ ,
}
and the BPS condition \zeroev\ becomes
\eqn\eombetwo{
\partial_m \sigma \partial_m \sigma  - e^{\sigma} \partial_m \chi
\partial_m \bar\chi - 
e^{-2\sigma}\partial_m H\partial_m H= 0 \ . 
}
These equations simplify due to the assumed symmetry.
By combining \zeroev\ and \eoma\  $\chi,\bar{\chi}$ can be eliminated and
one  obtains an ordinary
differential equation for $\sigma$ alone:
\eqn\eqb{
\sigma^{\prime\prime} + {4\over r}\sigma^\prime + {1\over 2}
(\sigma^\prime)^2 + 
{9\over 2}{Q_5^2\over (8\pi^2)^2} {1\over r^8} e^{-2\sigma} = 0 \ ,
}
where the prime denotes differentiation with respect to $r$.
It is straightfoward to check that 
\eqn\fsol{
\sigma(r) = \sigma_\infty+\ln(1 + {b_1 \over r^3})+\ln (1 + {b_2 \over r^3})
}
is the most general solution of  \eqb,  provided that the following
relation between the parameters $\sigma_\infty,b_1,b_2$ and $Q_5$ holds: 
\eqn\conda{
(b_1-b_2)^2 - {Q_5^2\over (8\pi^2)^2} e^{-2\sigma_\infty}  = 0 \ .
}

The equations of motion for $\chi$ and $\bar\chi$ following from
variations of \fivdace\ which vanish at the boundary \foot{Because some
fields are singluar at $x=x_0$, one has to exclude an 
infinitesimal sphere around $x_0$, in addition to the boundary at
infinity.} read  
\eqn\chieqa{
\eqalign{\partial_n\big(e^\sigma \partial_n \chi- \partial_n H
\chi\big)&=0 \ , \cr
\partial_n\big(e^\sigma \partial_n \bar\chi+ \partial_n H
\bar\chi\big)&=0 \ . }
} 
Due to radial symmetry these equations can be integrated once and
one gets two first order differential equations 
\eqn\eqd{
\eqalign{e^\sigma \chi^\prime +{Q_5\over 8\pi^2}{1\over r^4} \chi &=
{\alpha Q_5\over8\pi^2  }{1\over r^4} \ ,\cr 
e^\sigma \bar{\chi}^\prime -{Q_5\over
8\pi^2}{1\over r^4} \bar{\chi} 
&= {\bar{\alpha}Q_5\over 8\pi^2}{1\over r^4}\ , \cr }
}
where $\alpha$, $\bar{\alpha}$ are integration constants.
Note that the charges \twobch\ are related to the
integration constants in 
\eqd\ by  $Q_2=\alpha Q_5,\bar Q_2=\bar\alpha Q_5$.

These equations are easily integrated, and the result is 
\eqn\chisol{\eqalign{
\chi(r) &= {q_2\over Q_5}\big({{1+b_1/r^3}\over{1+b_2/r^3}}\big) +
\alpha \ , \cr
\bar{\chi}(r) &= -{\bar q_2\over Q_5}
\big({{1+b_2/r^3}\over{1+b_1/r^3}}\big) + \bar\alpha \ . }
}
The integration constants
$\alpha$, $\bar\alpha$ can be related to the asymptotic values
of the fields $\chi$, $\bar\chi$ at infinity: 
\eqn\asychi{
\eqalign{
\chi_\infty &=  \alpha +{q_2\over Q_5},\cr
\bar\chi_\infty &= \bar\alpha- {\bar q_2\over Q_5} \ . }
}

So far only one linear combination of equations \zeroev\ and \eoma\ has
been used. Requiring that these are both satisfied in addition
to \conda\ imposes a second condition on the parameters:
\eqn\condb{
(b_1+b_2)^2-  {Q_5^2\over (8\pi^2)^2} e^{-2\sigma_\infty} - 
{q_2\bar q_2 \over (8\pi^2)^2} e^{-\sigma_\infty}  =0 \ .
}
The two relations \conda\ and \condb\ determine the parameters $b_1$,
$b_2$ in \fsol\ in terms of the asymptotic value of $\sigma$ and the
charges,
\eqn\bonebtwo{\eqalign{b_1&= {1\over 16 \pi^2}\Big( \sqrt{Q_5^2
e^{-2\sigma_\infty} +  
q_2\bar q_2  e^{-\sigma_\infty}}+ Q_5e^{-\sigma_\infty}\Big),\cr
b_2&= {1\over 16 \pi^2}\Big( \sqrt{Q_5^2 e^{-2\sigma_\infty} + 
q_2\bar q_2  e^{-\sigma_\infty}}-Q_5e^{-\sigma_\infty}\Big).}}
Exchanging $b_1$ and $b_2$ in the solution, corresponds to exchanging 
$\chi_1$ and $\chi_2$.
In the following section it is shown that the instanton action is
independent of the asymptotic values of $\chi$, $\bar\chi$ at infinity. 

Note that choosing the five dimensional metric to be flat is 
consistent, 
since the stress tensor derived from \fivdace\ vanishes on shell.

In general the instanton solution constructed above 
carries three charges
$Q_5,q_2,\bar q_2$ and naturally  generalizes the solutions carrying only
a single charge given in \beckera\beckerb\lusta. The Euclidean instanton
solutions have an interpretation as (complex) saddle points of the Euclidean
action which
describe tunneling between vacua with different charges. Note that away 
from asymptotic infinity the solution becomes complex, since \chisol\
implies that $\chi$ is the complex conjugate of $\bar \chi$ only
in the limit  $r\to \infty$.

\newsec{The instanton action}

The Euclidean action \fivdace\ is not invariant under constant shifts
of $\chi,\bar\chi$. The invariance under such shifts can be
restored by adding a total derivative term to the action \fivdace: 
\eqn\stileq{
S^\prime_{eucl}=S_{eucl}+ {1\over 4} \int
d^5x\sqrt{g}\nabla^n\Big( e^{\sigma}
(\nabla_n\chi\bar\chi+\nabla_n\bar\chi\chi)\Big) \ .
}
Under constant shifts of $\chi,\bar\chi$ the action ${S^\prime}$
transforms into itself up to a 
term proportional to the equation of motion.

Adding  a
total derivative term  is equivalent to adding a boundary term to the
action and this does not change the
equations of motion. The boundary term does however contribute to the
instanton action.
Using the solutions of the equations of motion in the action \stileq\  
gives 
\eqn\instacf{
\eqalign{S_{inst}&=- \oint \partial_n \sigma }  
}
(the equation of motion \eombetwo\ was used to express the action as
a total derivative). Evaluating \instacf\ using the solution for $\sigma$
given in \fsol\ 
together with \bonebtwo\ leads to 
\eqn\instacg{
S_{inst}= 8\pi^2(b_1+b_2)=\sqrt{Q_5^2e^{-2\sigma_\infty}+
q_2\bar q_2 e^{-\sigma_\infty}} \ .
}
Note that, as expected, the boundary term in \stileq\ removes the
dependence of the instanton action on $\alpha$, $\bar\alpha$ (or
equivalently on $\chi_\infty, \bar\chi_\infty$). The
weight with which an instanton contributes in physical processes is given
by 
$\exp(-S_{inst})$. 
Just as in the case of the $\theta$ term in the action for Yang-Mills
instantons the only dependence on the asympotic value of the 
axionic scalars $D,\chi,\bar\chi$ is given by a phase factor
$\exp(i\theta)$ 
\eqn\phasent{
\exp(i\theta)=\exp\Big(i (Q_5 D_\infty+{1\over 4}Q_2
\bar\chi_\infty+ {1\over 4}\bar Q_2 \chi_\infty)\Big).
}
This term will imply that only discrete shifts of the axionic scalars
are a symmetry when instantons are taken into account \beckerb \foot{The discussion of the phase is not complete without
the investigation of the one loop determinants, but this lies beyond the
scope of our paper.}. In the path
integral this term has the role of a boundary term imposes  fixed boundary
conditions on the  charges \coleman\stromingerc.

This form of the instanton action generalizes the instanton actions for a
single charge. When two of the three charges $Q_5$, $Re(q_2), Im(q_2)$
are  set to zero it reduces to the
familiar expression for the fivebrane and membrane instanton actions. 

Reduction of the five dimensional theory on a circle relates M-theory
compactified on a Calabi-Yau manifold to Type IIA string theory
compactified on the same Calabi-Yau. The 
five-dimensional universal hypermultiplet turns into the four dimensional 
one. The Calabi-Yau volume (breathing mode) $\sigma$ is mapped  to the four  
dimensional dilaton 
by $\sigma=2\phi_{IIA}$ \stromingera\beckerb. The four dimensional
instanton 
action is then obtained by this replacement. The resulting dependence on
the string coupling $g=e^{\phi_{IIA}}$ is 
what is expected for NS-fivebrane and D2 brane instantons respectively.

\newsec{Broken and unbroken supersymmetries}

The explicit form of the unbroken supersymmetries can be determined from
the conditions \susytrafh,\susytrafg. Substituting the condition
$\delta\xi^i=0$ into $\delta\psi^i=0$ gives the following equations for the
supersymmetry transformation parameters $\epsilon^i$ :
\eqn\susydiffa{
\eqalign{(\nabla_r + {1\over 3} \partial_r \sigma +{1\over
4} e^{-\sigma} \partial_r
H)\epsilon^1&=0 \ , \cr 
(\nabla_r + {1\over 3} \partial_r \sigma -{1\over 4}e^{-\sigma} \partial_r
H)\epsilon^2&=0 \ .}
}
Using the explicit form of $\sigma$ given in \fsol\ and $H$ given in \etoc\
one finds  
\eqn\fermsol{
\eqalign{\epsilon^1(r) &= \Big(1+{b_1\over
r^3}\Big)^{-7/12}\Big(1+{b_2\over r^3}\Big)^{-{1/12}}\hat\epsilon^1,\cr
 \epsilon^2(r) &= \Big(1+{b_2\over
r^3}\Big)^{-{7/ 12}}\Big(1+{b_1\over r^3}\Big)^{-{1/
12}}\hat\epsilon^2 \ ,}
}
where $\hat \epsilon_1,\hat\epsilon_2$ are constant spinors. The four
unbroken supersymmetries are then found by demanding that
$\delta\xi_i=0$. Using \fermsol\ this condition reduces to a matrix
equation for the $\hat\epsilon_i$.
\eqn\matrixew{
\pmatrix{b_1+b_2+e^{-\sigma_\infty}{Q_5\over
8\pi^2}&ie^{-{\sigma_\infty/ 2}}{q_2\over 8\pi^2}\cr
-ie^{-\sigma_\infty/2}{\bar q_2\over 8\pi^2}&
b_1+b_2-e^{-\sigma_\infty}{Q_5\over 
8\pi^2}}\pmatrix{\hat\epsilon^1\cr
\hat\epsilon^2}=0
}
Note that the condition \condb\ is the same as the BPS condition that the  
determinant of the matrix \matrixew\ vanishes.
 
It is well known that the broken supersymmetries in the instanton
background will generate fermionic zero modes. Scattering amplitudes
vanish in the instanton background unless the fermionic zero modes are
absorbed by field insertions. This leads to new instanton induced
interactions. The simplest of such interactions are four fermion terms
\beckera\beckerb, which involve the Riemann tensor of the quaternionic
moduli space.  The presence of such terms  presumably implies 
that the metric on the quaternionic moduli also receives instanton
corrections.

The explicit form of the four-fermion terms will not be discussed 
here, but it is easy to  
see that the fermionic zero modes will be normalizable. 
As defined in \susytrafh\ the hyperinos transform as $\delta
\xi^i=M^i_{\;j}\epsilon^j$. In the instanton background one has
$\det(M)=0$, 
whereas the trace of the matrix $M$ is given by $2/3\gamma^r\partial_r
\sigma$. Hence the broken supersymmetries will correspond to the eigenvector
with eigenvalue $2/3\partial_r
\sigma$. The norm of the broken supersymmetry $|\delta\xi|^2$ will
therefore contain 
\eqn\sqrint{
\partial_r\sigma\partial_r\sigma= {1\over r^8}{(b_1+b_2+2b_1b_2/r^3)^2\over 
(1+b_1/r^3)^2(1+b_2/r^3)^2}
}
This expression 
behaves like $1/r^8$ as $r\to \infty$ and like  $ 1/r^2$ as $r\to 0$. Hence
$|\delta \xi|^2/|\epsilon|^2$   has a finite integral over $R^5$ and the
broken supersymmetries will be normalizable. In principle one can use the
explicit form of the solution to calculate the form of the instanton induced
terms.

\newsec{Discussion}

This paper presented a solution of the Euclidean equations of motion for
the universal hypermultiplet in five dimensional $N=2$ supergravity, which
generalizes solutions which carry only one charge (and reduces to them in
the limit when two of the three charges are set to zero).  The action of
the instanton has interesting properties.  The action of a multiple charge
instanton is not the sum of the actions of instantons carrying a single
charge: its form is characteristic of a ``non threshold bound state''. This
means that the instantons cannot be separated: a configuration of
single-charge instantons at different spacetime points would not be
supersymmetric. Since the fivebrane and two brane solutions preserve
different supersymmetries this has to be the case.  In
\towna\townb\ a solution of eleven dimensional supergravity in flat space
was found which can be interpreted as a non threshold bound state of a M2
brane within a M5 brane. In \townd\ it was shown that the tensions behave
exactly as in \instacg.

It would be very interesting to understand the geometric conditions for the
supersymmetric wrapping of a Euclidean M5 brane on a Calabi-Yau which
follow from the existence of BPS instantons in analogy with the analysis in
\beckera. Presumably one would need to understand the Euclidean fivebrane
worldvolume theory \pasti\schwarzd\ in the presence of three form fluxes
\westb\lustc. Since the 
worldvolume action of M5 branes is considerably more complicated than the
one of M2 branes this is not straightforward (see however \stromingerd,
where the M5 brane worldvolume theory is related to Kodaira-Spencer
theory).  The use of 
calibrations might also prove useful in this respect.

The classical saddle point of the Euclidean action discussed here 
is only the starting point of an instanton calculation. In
particular, integration over fermionic zero modes and the evaluation of
fluctuation determinants would have to be performed to obtain an explicit
form of the corrections to the hypermultiplet moduli space metric.
Fermionic zero modes can
be analyzed to some extent within the framework of  supergravity, as
briefly indicated in this note. 
However, as discussed in
\harveyb, the rules for calculating the fluctuation determinants are
not properly understood M-theory, unlike in field theories where a second
quantized path integral is at our disposal. This is obviously an important
open problem. As is sometimes the case with D-instantons, the best hope
might be to use heterotic -- type II duality, which relates spacetime
instantons of type II string theory to world sheet instantons on the
heterotic side, to learn something about instanton calculus for $N=2$
backgrounds.

\medskip
{\bf Acknowledgements}
\medskip

We are grateful to A. Strominger and V. Balasubramanian for useful
conversations.
The work of M.G. is supported in part by the David and Lucile Packard
Foundation. The work of M.S. was supported in part by NSF grant
PHY-98-02709.

\listrefs
 
\end